\documentstyle[preprint,aps,epsfig]{revtex}

\newcommand{\be}{\begin{equation}}
\newcommand{\ee}{\end{equation}}
\newcommand{\Dlt}{\Delta}
\newcommand{\dlt}{\delta}
\newcommand{\om}{\omega}

\newcommand{\br}{{\bf r}}
\newcommand{\vp}{\varphi}
\newcommand{\ep}{\varepsilon}
\newcommand{\bt}{\beta}
\newcommand{\al}{\alpha}
\newcommand{\gm}{\gamma}

\newcommand{\ra}{\rightarrow}

\newcommand{\prt}{\partial}
\newcommand{\dgr}{\dagger}

\tightenlines

\begin{document}

\draft

\title{Regulating entanglement production in multitrap 
Bose-Einstein condensates} 
\author{V.I. Yukalov$^{1,2}$ and E.P. Yukalova$^{1,3}$} 

\address{$^1$Institut f\"ur Theoretische Physik, \\
Freie Universit\"at Berlin, Arnimallee 14, D-14195 Berlin, Germany}
\address{$^2$Bogolubov Laboratory of Theoretical Physics, \\
Joint Institute for Nuclear Research, Dubna 141980, Russia}
\address{$^3$Department of Computational Physics, Laboratory of Information
Technologies, \\
Joint Institute for Nuclear Research, Dubna 141980, Russia}

\maketitle

\vskip 2cm

\begin{abstract}

A system of traps is considered, each containing a large number of 
Bose-condensed atoms. This ensemble of traps is subject to the action 
of an external modulating field generating nonequilibrium nonground-state 
condensates. When the frequency of the modulating field is in resonance 
with the transition frequency between two different topological coherent
modes, each trap becomes an analog of a finite-level resonant atom. 
Similarly to the case of atoms in an electromagnetic resonant field, 
one can create entanglement between atomic traps subject to a common 
resonant modulating field generating higher coherent modes in each of 
the traps. A method is suggested for regulating entanglement production 
in such a system of multitrap and multimode Bose-Einstein condensates 
coupled through a common resonant modulating field. Several regimes 
of evolutional entanglement production, regulated by manipulating the 
external field, are illustrated by numerical calculations.

\end{abstract}

\vskip 1cm


\newpage

\section{Introduction}

Entanglement is a notion that is important for quantum computation and 
quantum information processing [1--5]. Considering entanglement, one usually 
studies the systems of resonant atoms, possessing internal states, or spin 
assemblies [6--8]. As possible candidates for engineering entanglement,
trapped ions have also been considered [1,2]. On the other hand, recent 
advances in atomic physics with neutral atoms, experiencing Bose-Einstein
condensation [9--14], have led to several suggestions of employing these 
condensed atoms for producing entanglement. Thus, entanglement was shown 
to be feasible in Bose-condensed systems of atoms with two different 
internal states or with two collective modes [15,16]. Entanglement can 
be realized between the atoms from two clouds with different momenta, 
which arise in the process of coherent scattering of light from a 
Bose-Einstein condensate [17--20]. It is also possible to consider 
entanglement in Fermi-Bose mixtures [21--23] and in atom-molecule mixtures 
created by Feshbach resonance [24,25]. A review of the general properties
of two or more entangled modes can be found in Ref. [26].

A promising and efficient way of generating several topological coherent 
modes in Bose-condensed systems is by means of the resonant modulation of
the external trapping potential [27--30]. The properties of these coherent
modes have also been studied in a number of works [31--45]. A Bose-condensed
system with so generated coherent modes possesses a rich behaviour displaying
several nontrivial effects, such as interference fringes [29,41], interference
current [29,41], mode locking [27--29], dynamic transitions and critical
phenomena [28,29,37], chaotic motion [30,44], harmonic generation and
parametric conversion [30,44], and atomic squeezing [29,43]. It was mentioned 
[42,43] that a trap with several generated coherent modes is an analog of a 
finite-level resonant atom.

In the present paper, we consider not just a single trap, but an ensemble 
of many traps subject to the action of a common resonant modulating field,
generating topological coherent modes in each of the traps. Such a system 
is equivalent to an ensemble of many resonant atoms subject to a common 
electromagnetic field. As is known [6], the atoms, resonantly interacting 
with a common field, become entangled, which is termed resonant entanglement. 
In the same way, there appears entanglement between the traps subject to 
a common resonant field generating coherent modes. This {\it resonant 
entanglement} is a particular case of entanglement between coherent modes 
[26]. One of possible ways of creating a multitrap Bose condensate could
be by means of optical lattices, which should be sufficiently deep, so that 
a potential well around each lattice site would contain a large number of 
atoms, thus, representing a separate trap. Then the lattice as a whole 
should be shaken by an external modulating field, resonantly generating 
coherent modes in each of the lattice-site wells. In an experiment, the
resonant field can be produced by a time-dependent modulation of magnetic 
fields or by laser beams [27--31]. The advantage of employing a deep optical 
lattice as a multitrap device is the possibility of varying the barriers 
between lattice sites and, hence, the feasibility of tuning the properties 
of the overall system.

A peculiar feature of entanglement, occurring in multitrap multimode Bose 
condensates, is that this is an entanglement between mesoscopic (or almost
macroscopic) objects, as compared to the entanglement between atoms, which 
are microscopic objects. At the same time, a mesoscopic Bose condensate 
is a quantum object, described by a wave function satisfying a nonlinear 
Schr\"odinger equation. This unique combination of being mesoscopic in 
size and quantum in its features makes Bose-Einstein condensate a specific
matter. A multitrap collection of multimode condensates constitutes a
novel system that can find applications in quantum computing and information
processing.

Before going to a mathematical description, let us present in more detail 
the physical picture we keep in mind. A set of Bose-Einstein condensates can 
be created by loading ultracold atoms into a deep optical lattice. Thus a 
one-dimensional array of $^{87}$Rb condensates was formed [46], with a period 
of $2.7\times 10^{-4}$ cm. The lattice was sufficiently deep, with the depth 
$V_0\approx 2\pi\times 50$ kHz $\approx 3\times 10^5$ Hz, or, in units of 
the recoil energy, $V_0/E_r\approx 600$. The total number of lattice sites 
was 30, each containing $\sim 10^4$ atoms. A similar experiment for $^{87}$Rb 
condensates was reported in Ref. [47]. A deep one-dimensional optical lattice,
with adjacent sites spaced by $5.3\times 10^{-4}$ cm, contained about $10^4$ 
atoms. The number of lattice sites could be varied between 5 to 35, so that 
the number of condensed atoms in each site was between 100 to $10^3$. In these
experiments [46,47], the number of condensed atoms $N$ in each of the lattice 
sites was very large, reaching $N\sim 10^4$. Each lattice site played the role
of a trap containing a large number of completely Bose-condensed atoms in a 
well defined coherent state. But, because of the great depth of each site, the
intersite tunneling was strongly suppressed. Such a state of matter is commonly
called the Mott insulator. A specific feature of the insulating phase, formed 
in the experiments [46,47], is an enormous filling factor, with the number of 
atoms per lattice site being very large, $N\gg 1$. These insulating systems 
were probed [46,47] by releasing the atoms from their lattice, so that the 
wave packets of different condensates, emanating from separate lattice sites, 
could expand and overlap. A surprising discovery of these observations was 
that, contrary to what could be naively expected, there existed high-contrast 
interference fringes, which is commonly associated with the presence of 
coherence between different lattice sites. The interference-fringe contrast 
was indeed comparable to the results observed in the early interference 
experiments for two well-correlated condensates (see discussion in the review 
papers [10,14]). Similar results were also obtained for deep optical lattices 
with low filling factors, showing that some kind of coherence persists even 
deep in the insulating phase, preserving a good visibility of the interference 
patterns observed after free expansion [48]. The theoretical explanation 
of these interference effects shows that, though separate lattice sites may 
look to a first glance independent, the overall insulating phase is a highly 
correlated state of the many-body system, being a generically entangled state 
[49].

An additional degree of entanglement for such deep lattices, with large 
filling factors, is provided by subjecting the whole system to the action 
of an external resonant field, being {\it common} for all lattice sites. 
Since the resonant field is common for the total lattice, one has to treat 
the insulator plus the field as a whole, as a single complex object. This 
situation is completely analogous to what one has subjecting an ensemble 
of finite-level atoms to a common resonant field, realizing the so-called 
resonant entanglement [6]. There is a straightforward and direct analogy 
between these two cases [41,42,50].

A Bose-condensed gas can also be made equivalent to a finite-level system. 
For this purpose an external alternating field, acting on the Bose-Einstein 
condensate of trapped atoms, has to oscillate with a frequency $\om$ close 
to the transition frequency $\om_0$ between the ground-state level and 
that of the chosen excited mode. Because of the nonlinearity of the stationary
Gross-Pitaevskii equation, the spectrum of self-consistent collective modes is
not equidistant. This spectrum, for different trap shapes, was calculated and
analysed in Refs. [10,27,34,38,39,42]. Applying several driving fields, with
the appropriately tuned frequencies, a multimode system can be formed. Such a
finite-level condensate is completely analogous to a finite-level atom subject
to the action of a resonant electromagnetic field [41,42]. In the process of
the resonant excitation of a Bose condensate, similar to the same procedure
for an atom, there exist as well nonresonant transitions. However, their
probability is much lower than that of the chosen resonant transition. The
effect of nonresonant transitions was studied in Ref. [29], where it was shown
that, for a typical setup, the occurrence of nonresonant transitions would
destroy the condensate during the time approximately equal to the lifetime of
atoms in a trap. Therefore, the resonant generation of nonground-state
modes in a Bose condensate is a feasible procedure. These conclusions were
obtained by employing accurate analytical calculations as well as by direct
numerical simulations for the temporal Gross-Pitaevskii equation [30,44,45].
The resonant generation of a dipole mode was realized in experiment [32].
Vortices are the examples of the topological modes, which can be generated
in the resonant way [31].

Each lattice site of a deep optical lattice represents a trap
that may contain a large number of Bose-condensed atoms. According to 
experiments [46,47], this number can be as large as $N\sim 10^4$. A
collection of the lattice sites composes a strongly correlated system. The
correlation between different sites comes from three sources. First, the
standard method of preparing the considered system is by initially condensing
a large atomic cloud and then imposing on it a periodic lattice formed by
laser beams. Hence, at the initial time all atoms compose a general condensed
system in a single coherent state, after which they are loaded in different
lattice sites [46--48]. Second, though the tunneling between deep lattice
sites is small, but, nevertheless, it is finite, because of which phase
coherence on short length scales persists even deep in the insulating phase
[48]. Finally, when the overall lattice is subject to a {\it common} resonant
field, the whole system of the lattice plus field must be treated as one
complex object.

We show below that not merely one is able to generate entanglement in a
multitrap Bose condensate, but, moreover, it is possible to effectively
regulate the amount of the produced entanglement by manipulating with the
applied resonant field. The potentialities of regulating the entanglement
generation is what makes the considered method important for
applications. Using numerical calculations, we present several regimes
of regulated entanglement production, which illustrate wide possibilities
of the suggested approach. In the last Section V, we discuss how the
entanglement in the system can manifest itself in experiment.

Throughout the paper, the system of units is employed, where the Planck
constant and Boltzmann constant are set to unity, $\hbar\equiv 1$, 
$k_B\equiv 1$.

\section{Single-Trap Coherent States}

Before studying a collection of many traps, let us, first, give an accurate
mathematical description of coherent states in a single trap. The intention
of the present section is to introduce the definitions and notations
characterizing a single trap, which will be used in the following section
for a generalization to the case of many traps. Let a trap contain $N$ Bose
atoms at a very low temperature, when the whole system is Bose-condensed.
The Bose-Einstein condensate, as is known [10,51], corresponds to a coherent
state $|\eta>$ of a Fock space. The coherent state is an eigenstate of a
destruction field operator, whose eigenvalue is given by a coherent field
$\eta(\br)$. In the Fock space, a normalized coherent state is represented
by a column
\be
\label{1}
|\eta>\; = \; \left [ \frac{e^{-N/2}}{\sqrt{n!}}\;
\prod_{i=1}^n \eta(\br_i) \right ]
\ee
with respect to the index $n=0,1,2,\ldots$, enumerating the column rows.
Equation (1) is the standard notation for a state in the Fock space [51].
A more detailed explanation of the notations and definitions, used in the 
paper, are given in Appendix. The scalar product of two coherent states 
is defined as
\be
\label{2}
<\eta_1|\eta_2>\; \equiv \; \sum_{n=0}^\infty \;
\frac{e^{-N}}{n!}\; (\eta_1,\eta_2)^n\; ,
\ee
where the scalar product of two coherent fields is
\be
\label{3}
(\eta_1,\eta_2) \equiv \int \eta_1^*(\br) \eta_2(\br)\; d\br \; .
\ee
The normalization conditions for the coherent states and fields are
\be
\label{4}
<\eta|\eta>\; =  \; 1 \; , \qquad (\eta,\eta)=N \; .
\ee

A coherent state is a ground state for a Bose-condensed system [10,51,52].
Here we keep in mind a dilute system of atoms confined by means of a
trapping potential  $U(\br)$ and interacting through the local potential
$$
\Phi(\br) = \Phi_0\dlt(\br) \; , \qquad
\Phi_0 \equiv 4\pi\; \frac{a_s}{m} \; ,
$$
in which $a_s$ is a scattering length and $m$, atomic mass. The stationary
coherent fields are defined by the Gross-Pitaevskii equation
\be
\label{5}
\hat H[\eta_k]\; \eta_k(\br) = E_k\; \eta_k(\br) \; ,
\ee
with the nonlinear Schr\"odinger Hamiltonian
\be
\label{6}
\hat H[\eta] = -\; \frac{\nabla^2}{2m} + U(\br) + \Phi_0|\eta|^2 \; .
\ee
The eigenproblem (5) yields a spectrum of the energies $E_k$, labelled
by a multi-index $k$. The lowest energy $E_0\equiv\min_k E_k$ corresponds
to the usual Bose-Einstein condensate, with the related coherent field
$\eta_0(\br)$ being just the condensate wave function. The higher energies
$E_k$ correspond to the nonground-state condensates, with the wave functions
$\eta_k(\br)$ called the topological coherent modes [27--30]. Each mode is
normalized as
\be
\label{7}
(\eta_k,\eta_k) = N \; .
\ee
The modes are named topological, since they describe topologically different
distributions of atoms $|\eta_k(\br)|^2$. Recall a vortex that is the most
known example of a topological mode.

For each coherent mode $\eta_k(\br)$, one may construct, similarly to Eq.
(1), a coherent state
\be
\label{8}
|\eta_k>\; \equiv \; \left [ \frac{e^{-N/2}}{\sqrt{n!}} \;
\prod_{i=1}^n \eta_k(\br_i)\right ] \; .
\ee
Equation (8), in the same way as Eq. (1), is the standard notation of a state
in the Fock space. The right-hand side here denotes a column with an infinite
number of rows enumerated by an index labelling the generic column element
inside the square brackets [51]. These mode coherent states are normalized to
one, $<\eta_k|\eta_k>=1$, by analogy with Eq. (4). The coherent modes of
Eq. (5), being defined by a nonlinear Schr\"odinger equation, are not
necessarily orthogonal. Therefore the coherent states (8), with a scalar product
\be
\label{9}
<\eta_k|\eta_p>\; = \; \exp\{ - N + (\eta_k,\eta_p)\} \; ,
\ee
look also nonorthogonal for $k\neq p$.

From the Cauchy-Schwartz inequality we know that
$$
|(\eta_k,\eta_p)|^2 \leq (\eta_k,\eta_k)(\eta_p,\eta_p) \; ,
$$
where the equality takes place if and only if $\eta_k(\br)=\eta_p(\br)$.
The labelling of the coherent modes is assumed to be done so that to avoid
the degeneracy in the notation. That is, $\eta_k(\br)$ can be equal to
$\eta_p(\br)$ then and only then, when $k$ coincides with $p$. Hence, for
different $k$ and $p$, one has
$$
|(\eta_k,\eta_p)| < N \qquad (k\neq p) \; .
$$
Therefore, in the limit of a large number of particles, for the scalar
product (9), we have
\be
\label{10}
\lim_{N\ra\infty} \; <\eta_k|\eta_p>\; = \; \dlt_{kp} \; .
\ee
This is equivalent to saying that the states (8) are asymptotically
orthogonal for $N\gg 1$. Recall that $N$ here is the number of atoms in
a single lattice site, which as is explained in the Introduction, can be
made in experiments very large.

It is worth stressing that the coherent states (8), composed of the
coherent fields $\eta_k(\br)$, characterize collective states of an
$N$-atomic system, and should not be confused with internal states of an
atom. To avoid such a confusion, we deal here with atoms possessing no
internal states.

\section{Multitrap Coherent States}

Now let us consider an ensemble of $L$ traps, each of which can possess
$M$ modes,
\be
\label{11}
L = \sum_j 1 \; , \qquad M= \sum_k 1 \; .
\ee
The traps are enumerated with an index $j=1,2,\ldots,L$. As is discussed
in the Introduction, such an ensemble of traps could be realized as an
optical lattice, with the potential wells at each lattice site deep enough
to incorporate a large number of atoms $N$. For sufficiently deep potential
wells, the condensate wave functions are localized close to the related
lattice sites [10,53], so that the coherent modes $\eta_{ik}(\br)$ and
$\eta_{jk}(\br)$ of different traps practically do not overlap,
\be
\label{12}
(\eta_{ik},\eta_{jk}) = \dlt_{ij} N \; .
\ee
This corresponds to an insulating lattice phase. For each coherent mode
$\eta_{ik}(\br)$, we construct a coherent state $|\eta_{ik}>$, by analogy
with Eq. (8). These coherent states, because of Eqs. (10) and (12), are
asymptotically orthogonal, such that
\be
\label{13}
<\eta_{ik}|\eta_{jp}>\; \simeq \dlt_{ij} \dlt_{kp} \qquad (N\gg 1) \; .
\ee
For each set $\{|\eta_{jk}>\}$ of the coherent states $|\eta_{jk}>$ a
resolution of unity
$$
\sum_k |\eta_{jk}><\eta_{jk}| \; \simeq \hat 1
$$
is asymptotically ($N\gg 1$) valid, understood in the weak sense [10,52].
Thus, the set $\{|\eta_{jk}>\}$ forms a basis, possibly overcomplete.

Let us define a Hilbert space
\be
\label{14}
{\cal H}_j \equiv \overline{\cal L}\{|\eta_{jk}>\}
\ee
as a closed liner envelope over the set $\{|\eta_{jk}>\}$. Then, ${\cal H}_j$
is a Hilbert space of the coherent states associated with a $j$-th trap. The
total Hilbert space for an ensemble of $L$ traps is the tensor product
\be
\label{15}
{\cal H} = \bigotimes_j {\cal H}_j \; ,
\ee
in which $j=1,2,\ldots,L$. The total space (15) contains entangled as well
as disentangled states. We define the set of disentangled states
\be
\label{16}
{\cal D} \equiv \{ \bigotimes_j |\vp_j>\; : \;  |\vp_j>\in
{\cal H}_j \}
\ee
as a subset of ${\cal H}$, composed of all disentangled states, having
the form of the tensor products $\bigotimes_j|\vp_j>$.

When we deal with an absolutely equilibrium system, then, as is evident,
all atoms condense to the lowest-energy state. This implies, assuming
that all traps are identical, that in each of them solely the states
$|\eta_{j0}>$ will be occupied, all other excited states being empty.
In order to physically realize a nontrivial situation, it is necessary
to invoke a mechanism for the creation of several coherent modes. Then,
clearly, we have to consider a nonequilibrium system. An efficient
mechanism for generating coherent modes has been suggested [27--30],
based on a resonant modulation of the trapping potential. Dealing with
a nonequilibrium system, the coherent states $|\eta(t)>$ from the Hilbert
space (15) become functions of time $t$, generally, having the form
\be
\label{17}
|\eta(t)>\; = \; \sum_k c_k(t) \bigotimes_j |\eta_{jk}> \; .
\ee
Here we assume that the same resonant pumping field acts on all traps,
generating the coherent modes with the probability $|c_k(t)|^2$.

Varying the coefficients $c_k(t)$ in Eq. (17), different entangled states
can be created. Maximal entanglement is achieved when $|c_k(t)|^2=1/M$. Thus,
for $|c_k|^2=1/2$, $L=2$ and $M=2$, Eq. (17) would represent a Bell state.
For $|c_k|^2=1/2$, $L>2$ and $M=2$, this would be a multicat state. And for
$|c_k|^2=1/M$, $L>2$, with $M>2$, a multicat multimode state would arise.

The statistical operator, corresponding to the state (17), is
\be
\label{18}
\hat\rho(t) = |\eta(t)><\eta(t)| \; .
\ee
To quantify the entanglement generated in the system, we shall employ the
measure of entanglement production introduced in Refs. [54,55]. The advantage
of this measure is its generality, which allows for the usage of the measure
for any multiparticle systems; for example, the four-particle entanglement
[56,57] can be easily characterized. The general measure of entanglement
production [54,55] can be defined for an arbitrary operator. Here we shall
use it for the statistical operator (18).

Let us define a nonentangling counterpart of the operator (18) as
\be
\label{19}
\hat\rho^\otimes(t) \equiv \bigotimes_j \hat\rho_j(t) \; ,
\ee
where the partite operators are
\be
\label{20}
\hat\rho_j(t) \equiv {\rm Tr}_{\{ {\cal H}_{i\neq j} \}} \hat\rho(t) \; .
\ee
Entanglement, generated by the statistical operator (18), is quantified
by the measure of entanglement production
\be
\label{21}
\ep(\hat\rho(t)) \equiv \log\;
\frac{||\hat\rho(t)||_{\cal D}}{||\hat\rho^\otimes(t)||_{\cal D}} \; ,
\ee
in which the logarithm is to the base 2 and
$$
||\hat\rho(t)||_{\cal D} \equiv \sup_{f\in{\cal D}} ||\hat\rho(t)f||
\qquad (||f||=1)
$$
is the operator spectral norm, associated with the corresponding vector
norm, and restricted to the set of disentangled states (16).

For what follows, we shall need the notation
\be
\label{22}
n_k(t) \equiv |c_k(t)|^2 \; ,
\ee
characterizing the mode probabilities, or the fractional mode populations.
As the partite operators (20), we get
$$
\hat\rho_j(t) = \sum_k n_k(t)|\eta_{jk}><\eta_{jk}| \; .
$$
Also, we find the norms
$$
||\hat\rho(t)||_{\cal D} = ||\hat\rho_j(t)||_{\cal D} =\sup_k n_k(t)\; ,
\qquad ||\hat\rho^\otimes(t)||_{\cal D} = [\sup_k n_k(t)]^L \; .
$$
Then the measure (21) of entanglement, generated by the statistical operator
(18), writes as
\be
\label{23}
\ep(\hat\rho(t)) = (1-L)\log\;\sup_k n_k(t) \; .
\ee
This measure, for the present case, has a nice property of being
expressed by the equation
\be
\label{24}
\ep(\hat\rho(t)) = (L-1)\ep_2(t)
\ee
through the bipartite measure
\be
\label{25}
\ep_2(t) \equiv -\log\; \sup_k n_k(t) \; .
\ee
The relation (24) here is valid for any number of traps $L$ and any number
of modes $M$. This relation is a particular concrete example of the
fundamental property, formulated by Linden, Popescu, and Wootters [58,59],
that "the parts determine the whole in a generic pure quantum state".

\section{Regulating Entanglement Production}

Entanglement is such a general effect that it exists practically for 
all condensed-matter systems, changing together with phase transitions
[54,55]. However, in the majority of cases, it is not easy to govern
the level of entanglement. To be practically useful, the system should
allow for an efficient way of manipulating with the generated entanglement,
as is the case of some simple spin systems [8].

Here we show how entanglement production can be regulated in a multitrap
Bose-condensed system. To excite the higher coherent modes, it is necessary
to use the method of resonant generation [27-30]. For this purpose, the
system is subject to the action of an external modulating field
\be
\label{26}
V(\br,t) = V_1(\br)\cos\om t + V_2(\br)\sin\om t \; ,
\ee
whose frequency is tuned close to the transition frequency
\be
\om_0 \equiv E_1 - E_0
\ee
between the ground-state mode, with an energy $E_0$, and an excited coherent
mode, with an energy denoted by $E_1$. The resonance condition
\be
\label{28}
\left | \frac{\Dlt\om}{\om}\right | \ll 1 \qquad (\Dlt\om \equiv \om -\om_0)
\ee
is assumed to be valid. For a nonequilibrium system, one has to consider
the time-dependent Gross-Pitaevskii equation
\be
\label{29}
i\; \frac{\prt}{\prt t}\; \vp(\br,t) = \left ( \hat H[\vp] +
\hat V \right )\; \vp(\br,t) \; ,
\ee
in which the wave function is normalized to one, $(\vp,\vp)=1$, the nonlinear
Schr\"odinger Hamiltonian is
\be
\label{30}
\hat H[\vp] = -\; \frac{\nabla^2}{2m} + U(\br) + \Phi_0 N |\vp|^2 \; ,
\ee
and $\hat V=V(\br,t)$ is the resonant field (26). We look for the solution
of Eq. (29) in the form
\be
\label{31}
\vp(\br,t) = \sum_k c_k(t) \vp_k(\br) e^{-iE_k t} \; ,
\ee
where the coefficients $c_k(t)$ change in time slower than the exponential
$\exp(-i E_kt)$. The quantity $|c_k(t)|^2$ defines the probability of
generating a $k$-th mode, or the fractional mode population. For these
probabilities, the normalization condition holds
\be
\label{32}
\sum_k |c_k(t)|^2  = 1 \; .
\ee
Since we have assumed that all traps are identical, and the same resonant
field acts on each of them, the probabilities $|c_k(t)|^2$ do not depend
on the trap index.

A resonant field of type (26) generates one higher coherent mode. Wishing
to create several topological coherent modes, one should apply several
resonant fields with the frequencies tuned close to the transition frequencies
of the chosen modes [30,44]. Here we consider a two-mode case, one of them
being the ground-state mode and another, an excited coherent mode.

Substituting the form (31) into Eq. (29) and employing the averaging
technique, we can derive the equations for the coefficients $c_k(t)$.
This derivation, with all details and with a full mathematical foundation
has been given earlier [27--30]. Below, we present only the results of this
procedure. We shall need the notation for the interaction transition
amplitude
\be
\label{33}
\al_{kp} \equiv \Phi_0 N\left ( |\vp_k|^2,\; 2|\vp_p|^2-|\vp_k|^2\right )
\ee
and the pumping-field transition amplitude
\be
\label{34}
\bt_{kp} \equiv (\vp_k,\hat B\vp_p) \; ,
\ee
where $\hat B\equiv V_1(\br)-i V_2(\br)$. For the two-mode case, it is
convenient to introduce the average interaction amplitude
\be
\label{35}
\al \equiv \frac{1}{2}\; \left ( \al_{01} + \al_{10}\right )
\ee
and to write the pumping-field amplitude in the form
\be
\label{36}
\bt_{01} = \bt e^{i\gm} \; , \qquad \bt\equiv |\bt_{01}| \; .
\ee
We denote the effective detuning as
\be
\label{37}
\dlt \equiv \Dlt + \frac{1}{2}\left (\al_{01} - \al_{10} \right ) \; ,
\ee
where
\be
\label{38}
\Dlt \equiv \Dlt\om + \al_{00} - \al_{11} \; .
\ee
The fractional mode amplitude can be written as
\be
\label{39}
c_k(t) = | c_k(t)| \exp\{ i \pi_k(t) t\} \; ,
\ee
where $\pi_k(t)$ is a real function representating the mode phase. The final
equations can be cast to the system of two real equations for the population
difference
\be
\label{40}
s(t) \equiv |c_1(t)|^2 - |c_0(t)|^2
\ee
and for the effective phase difference
\be
\label{41}
x(t) \equiv \pi_1(t) - \pi_0(t) + \gm + \Dlt \; .
\ee
These equations are
$$
\frac{ds}{dt} = -\bt\sqrt{1-s^2}\; \sin x \; ,
$$
\be
\label{42}
\frac{dx}{dt} = \al s + \frac{\bt s}{\sqrt{1-s^2}}\; \cos x + \dlt \; .
\ee
By scaling the time variable with $\al$, one can see that the behaviour
of the solutions to Eqs. (42) is defined by two dimensionless parameters
\be
\label{43}
b \equiv \frac{\bt}{\al} \; , \qquad \ep \equiv \frac{\dlt}{\al} \; .
\ee
The former parameter characterizes the strength of the pumping field, and
the second parameter describes an effective detuning from the resonance,
which can always be made small, $|\ep|\ll 1$. By solving Eqs. (42), one
finds the fractional mode populations
\be
\label{44}
n_0(t) = \frac{1-s(t)}{2} \; , \qquad n_1(t) = \frac{1+s(t)}{2} \; .
\ee
Finally, the entanglement generation is quantified by the measure (25)
which, for the considered two-mode case, is
\be
\label{45}
\ep_2(t) = -\log_2 \sup\{n_0(t), n_1(t)\} \; .
\ee
This measure changes with time in the interval $0\leq\ep_2(t)\leq 1$,
being maximal when $n_0=n_1=1/2$.

Solving Eqs. (42), we suppose that at the initial time $t=0$ solely the
ground-state mode exists, so that $s(0)=-1$ and $x(0)=0$. Hence, initially,
$n_0(0)=1$ and $n_1(0)=0$, because of which $\ep_2(0)=0$. Then the pumping
resonant field is switched on, which generates an excited coherent mode
and produces entanglement. There are two ways of governing the amount of
entanglement production.

First, one can vary the amplitude and frequency of the resonant pumping
field, choosing by this the required parameters (43), whose values define
two main regimes of an oscillatory behaviour of the population difference
$s(t)$, and so, the regimes of $\ep_2(t)$. These are the mode-locked and
mode-unlocked regimes [27--30,34,37,41--44]. We solve numerically Eqs. (42)
and calculate the measure of entanglement production (45). Keeping in mind
that the detuning can always be made small, we set $\dlt=0$. Then the value
$b_c=0.497764$ is the critical point for the change of the dynamical regimes.
Below $b_c$, the measure (45) oscillates with time, never reaching one. When
the parameter $b$, defined in Eq. (43), equals the critical point $b=b_c$,
then the oscillating $\ep_2(t)$ reaches one. For the dimensionless amplitude
of the pumping field $b>b_c$, the oscillations of $\ep_2(t)$ are always in
the interval between 0 and 1. But the oscillation period sensitively depends
on the value of $b$. Thus, for $b=0.5$, just a little above $b_c$, the period
of oscillations is more than twice shorter than that for $b=b_c$. The period
for $b=0.7$ is about eight times shorter than for $b=b_c$. Thus, by varying
the amplitude of the pumping resonant field, we can strongly influence the
evolution of $\ep_2(t)$ both in its amplitude and period of oscillations.

There is one more very interesting way of regulating entanglement generation,
which can be done by switching on and off the applied resonant field. Recall
that this alternating field can be easily produced by modulating the magnetic
field forming the trapping potential in magnetic traps or by varying the
laser intensity in optical traps. Then, it is possible to create various
sequences of pulses for $\ep_2(t)$. For example, by switching on and off
the resonant field in a periodic manner, we may form equidistant pulses
of $\ep_2(t)$, with all pulses having the same shape, as is demonstrated
in Fig. 1. But we can also switch on and off the pumping field at different
time intervals, thus, forming nonequidistant pulses, as is shown in Fig. 2.
The possibility of creating very different pulses is illustrated in Fig. 3.
Regulating entanglement production by means of a manipulation with the
resonant pumping field, it is feasible to organize a kind of the Morse
alphabet.

\section{Relation to Time-of-Flight Experiments}

It is important to add an idea how the entanglement generated in the system
could manifest itself in order to verify it in a measurement. This can be
done by observing the expansion of atomic clouds released from the optical
lattice. The absorption images and interference fringes, observed in such
time-of-flight experiments, are known to strongly depend on the initial
conditions of the atoms before their release. If the initial spatial
distribution of atoms in one case is essentially different from that in
another case, the interference pictures for such two cases will be noticeably
different. The generation of topological coherent modes, from one side, is
directly related to the level of the induced entanglement and, from another
side, drastically changes the spatial distribution of atoms. Recall that the
considered modes are termed topological exactly because of their principally
distinct from each other spatial shapes [27--30]. Radically differing initial
conditions, due to differently prepared modes, will inevitably result in
distinguishable one from another absorption images. As is seen from Eq. (17),
if there exists just one mode, say the ground-state mode, so that
$n_0=|c_k|^2=1$, hence the fractional populations for all other modes are
zero, then $|\eta(t)>$ in Eq. (17) is a product state, with no entanglement.
Respectively, the measures (24), (25), and (45) are identically zero. By
generating excited coherent modes, one makes the state (17) entangled. For
instance, in the case of two modes, the maximal entanglement is achieved
for $|c_k|=1/\sqrt{2}$, when Eq. (17) represents a multicat state. Then
the measure of entanglement production (24) reaches its maximal value
$(L-1)\log 2=L-1$, assuming that the logarithm is to the base two. Then
the measures (25) and (45) are also maximal, being equal to one.

Suppose that at the time $t_0$ atoms were released from their traps. In
the time-of-flight experiments one measures the atomic distribution
\be
\label{46}
\tilde\rho({\bf p}) \equiv \int \rho(\br,\br',t_0) \;
e^{-i{\bf p}\cdot(\br-\br')}\; d\br\; d\br' \; ,
\ee
in which ${\bf p}=m\br/(t-t_0)$ and $\rho(\br,\br',t_0)$ is the first-order
density matrix of atoms just before their expansion. The distribution (46) is
normalized to the total number of particles $NL$, with $N$ being the number
of atoms in each lattice site and $L$, the number of sites, so that
\be
\label{47}
\int\tilde\rho({\bf p})\; \frac{d{\bf p}}{(2\pi)^3} =
\int \rho(\br,t_0)\; d\br = NL \; ,
\ee
where $\rho(\br,t_0)=\rho(\br,\br,t_0)$ is the density of atoms before
their release.

The field operator, acting on the total Hilbert space (15), can be
represented as a direct sum
\be
\label{48}
\psi(\br) = \oplus_j \psi_j(\br)
\ee
of the operators $\psi_j(\br)$ acting on the spaces ${\cal H}_j$. By the
definition of the coherent states, one has
\be
\label{49}
\psi_j(\br)|\eta_{jk}>\; = \eta_{jk}(\br)|\eta_{jk} >\; .
\ee
For the atomic system in the state (17), the density matrix at the moment
$t_0$ is
\be
\label{50}
\rho(\br,\br',t_0) = \; <\eta(t_0)|\psi^\dgr(\br')\psi(\br)|
\eta(t_0)> \; .
\ee
From here, using Eqs. (48) and (49), we find
\be
\label{51}
\rho(\br,\br',t_0) = \sum_k \; \sum_{ij} \;
|c_k(t_0)|^2 \eta_{ik}(\br)\; \eta^*_{jk}(\br') \; .
\ee
Substituting this into Eq. (46), we get
\be
\label{52}
\tilde\rho({\bf p}) = \sum_k \; \sum_{ij} \; n_k(t_0)\;
\tilde\eta^*_{ik}({\bf p})\; \tilde\eta_{jk}({\bf p}) \; ,
\ee
were the Fourier transform
\be
\label{53}
\tilde\eta_{jk}({\bf p}) \equiv \int \eta_{jk}(\br) \;
e^{-i{\bf p}\cdot\br} \; d\br
\ee
is introduced. Defining also the partial distribution
\be
\label{54}
\tilde\rho_k({\bf p}) \equiv \sum_{ij}\; \tilde\eta^*_{ik}({\bf p})\;
\tilde\eta_{jk}({\bf p}) \; ,
\ee
we finally obtain
\be
\label{55}
\tilde\rho({\bf p}) = \sum_k \; n_k(t_0)\; \tilde\rho_k({\bf p})\; .
\ee
When there is a sole mode, say when $n_0(t_0)=1$, then there is no entanglement,
and the measured distribution of atoms in the time-of-flight experiment is
$\tilde\rho_0({\bf p})$. The generation of excited modes creates entanglement
and at the same time essentially changes the distribution of expanding
atoms (55).

To give a more explicit illustration, let us consider a one-dimensional array
of lattice sites in the tight-binding approximation. The ground-state wave
function of atoms in a $j$-th site can be described by the localized orbital
\be
\label{56}
\eta_{j0}(z) = \frac{1}{\pi^{1/4}}\; \sqrt{\frac{N}{l_0}}\;
\exp\left\{ -\; \frac{1}{2}\left ( \frac{z-a_j}{l_0}\right )^2\right \}\; ,
\ee
in which $a_j\equiv ja$, with $a$ being a lattice spacing and $j=0,\pm 1,
\pm 2,\ldots$.  The effective width of the atomic cloud, given by $l_0$,
is generally defined through a variational procedure taking into account
atomic interactions [10,27,42]. The Fourier transform of Eq. (56) is
\be
\label{57}
\tilde\eta_{j0}(p) = (4\pi)^{1/4}\sqrt{Nl_0}\; \exp\left ( -\;
\frac{1}{2}\; p^2l_0^2 -ipa_j\right ) \; .
\ee
For the corresponding distribution (54), we get
\be
\label{58}
\tilde\rho_0(p) = 2\sqrt{\pi}\; NL l_0 \exp\left (-p^2 l_0^2\right )
\left ( 1 + \frac{2}{L} \; \sum_{i< j} \; \cos pa_{ij}\right ) \; ,
\ee
where $a_{ij}\equiv a_i-a_j$. This is what one would observe in the
time-of-flight experiment, when all atoms before the release are in the
ground state and there is no entanglement.

Suppose now that, in addition to the ground-state mode, the first excited
mode is generated having in a $j$-th site the wave function
\be
\label{59}
\eta_{j1}(z) = \frac{1}{\pi^{1/4}} \; \sqrt{\frac{2N}{l_1}}\left (
\frac{z-a_j}{l_1}\right ) \exp\left \{ -\; \frac{1}{2}\left (
\frac{z-a_j}{l_1}\right )^2 \right \} \; .
\ee
Here $l_1$ is an effective width, which can be defined through a variational
procedure [10,27,42]. Calculating the Fourieir transform of Eq. (59),
we obtain
\be
\label{60}
\tilde\eta_{j1}(p) = - i2\pi^{1/4}\sqrt{Nl_1}\; pl_1 \exp\left ( -\;
\frac{1}{2}\; p^2 l_1^2 - ipa_j\right ) \; .
\ee
Consequently, the related atomic distribution is
\be
\label{61}
\tilde\rho_1(p) =  4\sqrt{\pi}\; NL p^2 l_1^2 \exp\left (-p^2l_1^2
\right )\left ( 1 +\frac{2}{L} \; \sum_{i<j}\; \cos pa_{ij}\right )\; .
\ee
The total observable atomic distribution in the two-mode case is the sum
\be
\label{62}
\tilde\rho(p) = n_0(t_0)\tilde\rho_0(p) + n_1(t_0)\tilde\rho_1(p) \; .
\ee
If both $n_0(t_0)$ as well as $n_1(t_0)$ are nonzero, then entanglement
is generated in the system. At the same time, the measurable distribution
(62) is essentially different from the situation when just one term
$\tilde\rho_0(p)$ is present. In the particular case of two sites, when
$L=2$, the terms in Eq. (62) are defined by
\be
\label{63}
\tilde\rho_0(p) =4\sqrt{\pi} \; Nl_0\exp\left (-p^2l_0^2\right )
(1 +\cos pa) \; ,
\ee
where $a$ is the intersite distance, and
\be
\label{64}
\tilde\rho_1(p) =8\sqrt{\pi} \; Np^2 l_1^3 \exp\left (-p^2l_1^2\right )
(1 +\cos pa) \; .
\ee
It is evident that the sum (62) of the distributions (63) and (64) is
drastically different from the sole term (63). This difference is especially
noticeable at the point $p=0$, where
\be
\label{65}
\tilde\rho(0) = n_0(t_0)\tilde\rho_0(0) =
8\sqrt{\pi}\; Nl_0 n_0(t_0) \; .
\ee
Value (65) diminishes as soon as the second mode is generated, so that
$n_0(t_0)$ becomes less than one.

In conclusion, a multitrap ensemble of multimode Bose-Einstein condensates,
subject to the action of a {\it common} resonant field, is analogous to
a system of finite-level atoms in a common resonant electromagnetic field.
A multitrap system can be formed, e.g., as an optical lattice with deep
potential wells, incorporating many atoms around each lattice site. In the
multitrap multimode condensate a high level of entanglement can be achieved.
By varying the amplitude and frequency of the pumping resonant field,
different regimes of evolutional entanglement can be realized. Moreover,
by switching on and off the pumping field in various ways, it is feasible
to create entanglement pulses of arbitrary length and composing arbitrary
sequences of {\it punctuated entanglement generation}. Such a high level
of admissible manipulation with and regulating of entanglement can, probably,
be useful for information processing and quantum computing.

\vskip 5mm

{\bf Acknowledgement}

\vskip 2mm

Financial support form the German Research Foundation is appreciated.

\newpage

{\Large{\bf Appendix. Representation of coherent states in the Fock 
space}}.

\vskip 5mm

In this Appendix, we give, according to the referee's advice, an explanation 
of the main notations and definitions, used in the paper, which are related 
to the representation of coherent states in the Fock space. We mention here 
only the basic facts. Substantially more details, with an accurate mathematical 
foundation, cane be found in the very good book by Berezin [51]. Of course, 
there exist many other publications expounding this material, for instance, 
the classical textbook by Schweber [60]. Our notations are close to those 
employed in these books [51,60].

Recall, first, that the Fock space is a direct sum
$$
{\cal F} = \oplus_{n=0}^\infty {\cal H}_n
$$
of $n$-particle Hilbert spaces ${\cal H}_n$ composed of symmetric or 
antisymmetric functions, according to whether the Bose-Einstein or Fermi-Dirac 
statistics are considered. For spinless bosons, treated in the paper, the wave 
function $f_n(\br_1,\ldots,\br_n)$ is symmetric with respect to the permutations
of the spatial variables $\br_i$, for all $i=1,2,\ldots,n$. The $n$-particle 
function $f_n(\br_1,\ldots,\br_n)$ pertains to the Hilbert space ${\cal H}_n$. 
A state $\vp$ of the Fock space ${\cal F}$ is represented by the column 
\begin{eqnarray}
\nonumber
\vp=\left [ \begin{array}{l}
f_0 \\
f_1(\br_1) \\
f_2(\br_1,\br_2) \\
f_3(\br_1,\br_2,\br_3) \\
\vdots \\
f_n(\br_1,\br_2,\ldots,\br_n) \\
\vdots \\
\end{array} \right ]
\end{eqnarray}
It is clear that to write down each time such a column would be a too 
cumbersome way. Therefore one commonly shortens the notation, representing 
the state $\vp\in{\cal F}$ as
$$
\vp = [f_n(\br_1,\br_2,\ldots,\br_n)] \; ,
$$
where only the generic element of the above column is written, with the 
index $n=0,1,2,\ldots$ enumerating the rows of the column. Thus, the index 
$n$ at the $n$-th row of the state $\vp$ tells that the $n$-particle wave 
function $f_n(\br_1,\br_2,\ldots,\br_n)$ is implied. The scalar product of 
two states $\vp,\vp'$ of the Fock space ${\cal F}$ is defined as
$$
(\vp,\vp') \equiv \sum_{n=0}^\infty (f_n,f_n') \; ,
$$
where
$$
(f_n,f_n') \equiv \int f_n^*(\br_1,\br_2,\ldots,\br_n) 
f_n'(\br_1,\br_2,\ldots,\br_n) \; d\br_1 \ldots d\br_n 
$$
is a scalar product in the $n$-particle Hilbert space ${\cal H}_n$. When a 
particular kind of wave functions $f_n(\br_1,\br_2,\ldots,\br_n)$ is kept in 
mind, this is often emphasized by denoting the corresponding state in the 
Fock space as
$$
\vp \equiv |f> \; = [f_n(\br_1,\br_2,\ldots,\br_n)] \; .
$$

Coherent states $|\eta>$ in the Fock space are defined as the eigenvectors 
of the annihilation field operator $\psi(\br)$, so that
$$
\psi(\br)|\eta> \; = \; \eta(\br)|\eta> \; ,
$$
where the coherent field $\eta(\br)$ plays the role of the eigenvalue. Being 
a member of the Fock space, a coherent state is also represented by a column
\begin{eqnarray}
\nonumber
|\eta> \; = \left [ \begin{array}{l}
\eta_0 \\
\eta_1(\br_1) \\
\eta_2(\br_1,\br_2) \\
\vdots \\
\eta_n(\br_1,\br_2,\ldots,\br_n) \\
\vdots \\
\end{array} \right ] \; ,
\end{eqnarray}
which is equivalently denoted by the accepted short-hand notation as
$$
|\eta> \; = \; [\eta_n(\br_1,\br_2,\ldots,\br_n)] \; .
$$
From the definition of the coherent states as the eigenvectors of the field 
operator $\psi(\br)$, and using the action of the latter yielding
$$
\psi(\br)|\eta> \; = \left [ \sqrt{n+1}\; 
\eta_n(\br_1,\br_2,\ldots,\br_n,\br)\right ] \; ,
$$
one gets the recursion relation
$$
\sqrt{n+1}\; \eta_{n+1}(\br_1,\br_2,\ldots,\br_n,\br) = \eta(\br)\;
\eta_n(\br_1,\br_2,\ldots,\br_n) \; .
$$
Solving this equation by repeated iterations, one finds
$$
\eta_n(\br_1,\br_2,\ldots,\br_n) = \frac{\eta_0}{\sqrt{n!}} \;
\prod_{i=1}^n \eta(\br_i) \; .
$$
The normalization condition $<\eta|\eta>=1$ gives
$$
|\eta_0| =\exp\left \{ -\; \frac{1}{2}(\eta,\eta)\right \} \; .
$$
Assuming that the coherent field $\eta(\br)$ is normalized to the number of 
particles $N$, one has
$$
(\eta,\eta) \equiv \int \eta^*(\br) \eta(\br)\; d\br =  N \; .
$$
In this way, for a coherent state $|\eta>$, up to a global phase factor, one
obtains
\begin{eqnarray}
\nonumber
|\eta> \; = \; e^{-N/2} \; \left [ \begin{array}{l}
1 \\
\eta(\br_1) \\
\frac{1}{\sqrt{2!}}\; \eta(\br_1)\eta(\br_2) \\
\frac{1}{\sqrt{3!}}\; \eta(\br_1)\eta(\br_2)\eta(\br_3) \\
\vdots \\
\frac{1}{\sqrt{n!}}\;\eta(\br_1)\eta(\br_2)\ldots\eta(\br_n) \\
\vdots \\
\end{array} \right ] \; .
\end{eqnarray}
This is exactly the coherent state (1) written down in the accepted 
short-hand notation.

As is seen, to define the whole coherent state $|\eta>$, one needs to find 
the coherent field $\eta(\br)$. The latter, for the considered case of a 
dilute Bose gas with contact interactions, is a solution of the 
Gross-Pitaevskii equation.

\newpage

\begin{center}
{\large{\bf Figure Captions}}
\end{center}

\vskip 2cm

{\bf Fig. 1}. Regulated equidistant pulses of $\ep_2(t)$, formed by
switching on and off the resonant field, with $b=0.7$, so that $\ep_2(t)$
equals one during the time intervals $\Dlt t=7.35$ (in units of $\al^{-1}$),
and it equals zero during the same intervals $\Dlt t=7.35$. Time is
measured in units of $\al^{-1}$.

\vskip 1cm

{\bf Fig. 2}. Nonequidistant pulses of $\ep_2(t)$, created by switching
on and off the pumping field, with $b=0.7$, at nonequal time intervals.
Time is measured in units of $\al^{-1}$.

\vskip 1cm

{\bf Fig. 3}. Regulated pulses of $\ep_2(t)$, for the same $b=0.7$, as
in Fig. 2, but for essentially different moments of switching on and off
the pumping field. Time is measured in units of $\al^{-1}$.

\newpage

\begin{figure}[h]
\centerline{\psfig{file=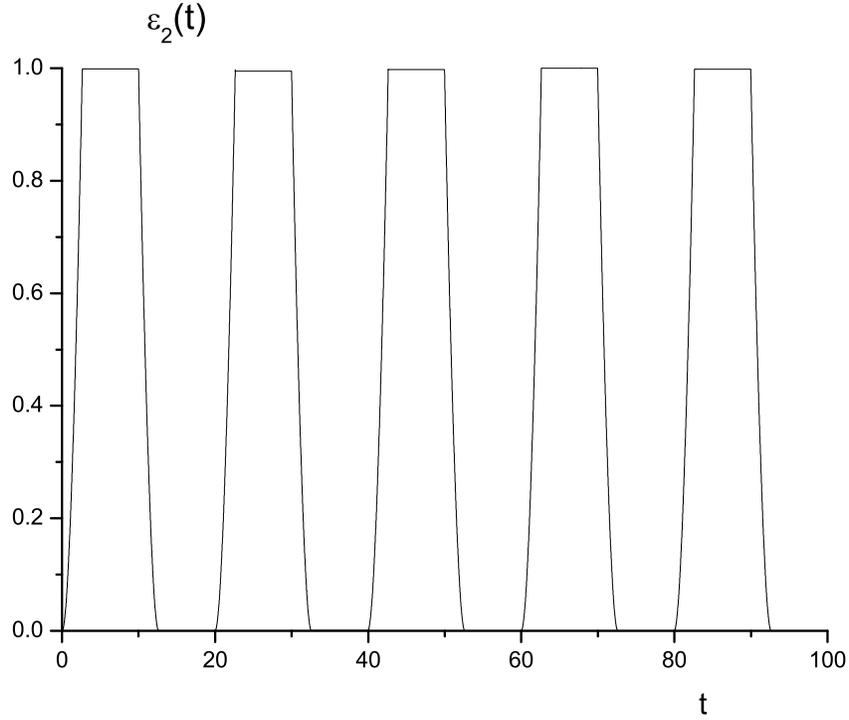,height=5in}}
\caption{Regulated equidistant pulses of $\ep_2(t)$, formed by switching
on and off the resonant field, with $b=0.7$, so that $\ep_2(t)$ equals one
during the time intervals $\Dlt t=7.35$ (in units of $\al^{-1}$), and it
equals zero during the same intervals $\Dlt t=7.35$. Time is
measured in units of $\al^{-1}$.}
\label{fig:Fig.1}
\end{figure}

\newpage

\begin{figure}[h]
\centerline{\psfig{file=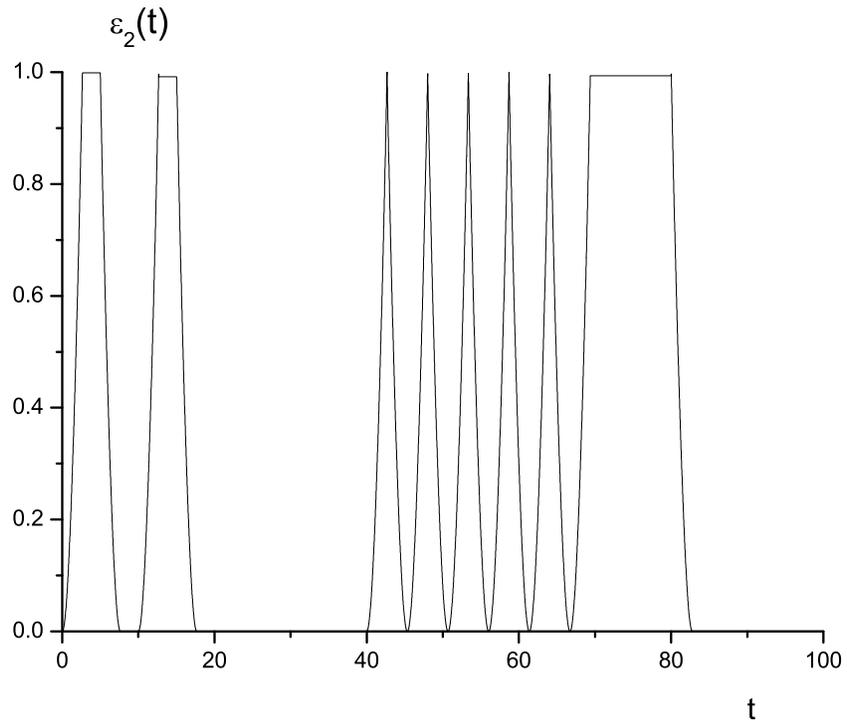,height=5in}}
\caption{Nonequidistant pulses of $\ep_2(t)$, created by switching on
and off the pumping field, with $b=0.7$, at nonequal time intervals.
Time is measured in units of $\al^{-1}$.}
\label{fig:Fig.2}
\end{figure}

\newpage

\begin{figure}[h]
\centerline{\psfig{file=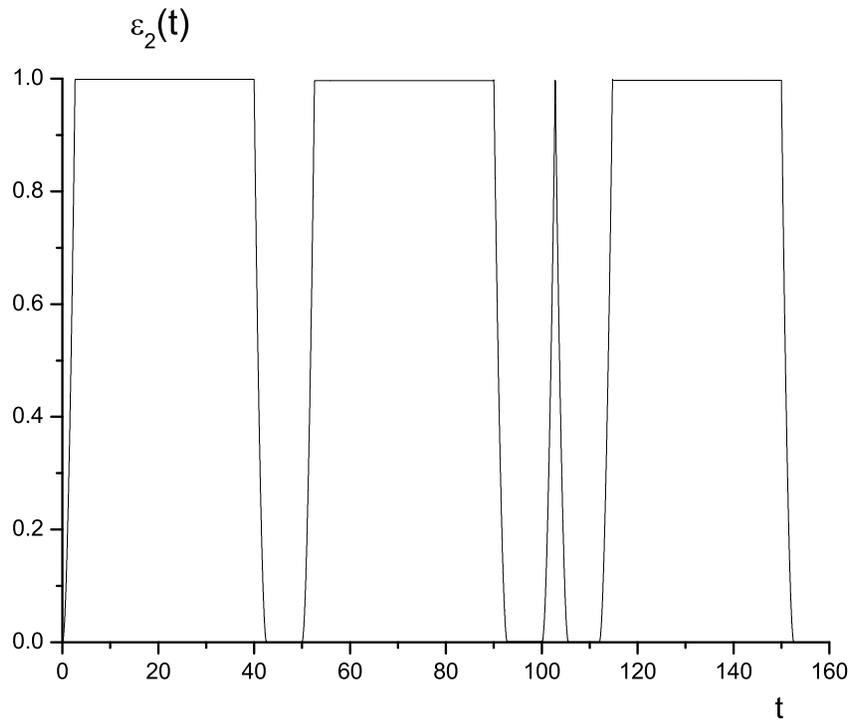,height=5in}}
\caption{Regulated pulses of $\ep_2(t)$, for the same $b=0.7$, as in
Fig. 2, but for essentially different moments of switching on and off
the pumping field. Time is measured in units of $\al^{-1}$.}
\label{fig:Fig.3}
\end{figure}

\end{document}